\documentclass[aps,russian,showpacs]{revtex4-1}
\usepackage[cp1251]{inputenc}
\usepackage[russian]{babel}
\usepackage[T2A]{fontenc}

\usepackage{graphicx,color}
\usepackage[unicode]{hyperref}

\begin{document}

\title{Исследование маятника Капицы.}

\author{Астрахарчик Г. Е.}
\affiliation{Departament de F\'{\i}sica i Enginyeria Nuclear, Campus Nord B4-B5, Universitat Polit\`ecnica de Catalunya, E-08034 Barcelona, Spain}

\author{Астрахарчик Н. А.}
\affiliation{142190 г. Троицк ул. Школьная 10-а, Лицей г. Троицка, Россия}

\begin{abstract}
Даже обыкновенный математический маятник обладает рядом интересных и фундоментальных свойств, поэтому он является популярным примером для применения как аналитических, так и численных методов. Еще более интересными свойствами обладает рассматриваемый ниже маятник Капицы. Для него выражение точного аналитического решения неизвестно, но возможно найти численное решение при помощи численного моделирования. Также полученные результаты можно сравнить с некоторыми предельными случаями, для которых известно решение.
\end{abstract}

\date{30 марта 2011 г.}

\maketitle

\section{Введение}

Маятником Капицы называется система, состоящая из грузика, прикрепленного к легкой нерастяжимой спице, которая крепится к вибрирующему подвесу. Маятник носит имя академика и нобелевского лауреата П.Л. Капицы, построившего в 1951 г. теорию для описания такой системы \cite{Kapitza1,Kapitza2}. В пределе неподвижной точки подвеса, модель переходит в обычный математический маятник, для которого имеются два положения равновесия: в нижней точке и в верхней точке. При этом равновесие математического маятника в верхней точке является неустойчивом и любое сколь угодно малое возмущение приводит к потере равновесия.

Удивительной особенностью маятника Капицы является то, что вопреки интуиции перевернутое (вертикальное) положение маятника может быть устойчивым в случае быстрых вибраций подвеса. Хотя такое наблюдение было сделано еще в 1908 г. А. Стефенсоном \cite{Stephenson}, в течении длительного времени не имелось математического объяснения причин такой устойчивости. П. Л. Капица подробно экспериментально исследовал такой маятник, а так же построил теорию динамической стабилизации, разделяя движение на <<быстрые>> и <<медленные>> переменные и введя эффективный потенциал \cite{Kapitza1,Kapitza2}. Инновационная работа П. Л. Капицы, опубликованная в 1951 году, открыла новое направление в физике --- вибрационную механику. Метод П.Л.Капицы используется для описания  колебательных процессов в атомной физике, физике плазмы, кибернетической физике.
Эффективный потенциал, описывающий <<медленную составляющую движения>>, описывается в томе <<механика>> курса теоретической физики Л.Д. Ландау\cite{LL}.

Маятник Капицы интересен еще и тем, что в такой простой системе можно наблюдать парметрические резонансы, когда нижнее положение равновесия не является больше устойчивым и амплитуда малых отклонений маятника нарастает со временем \cite{Butikov}. Также, при большой амплитуде вынуждающих колебаний в системе могут реализовываться хаотические режимы, когда в фазовой картине системы появляются странные аттракторы.

\section{Обозначения.}

Направим ось $y$ вертикально вверх, а ось $x$ горизонтально, так чтобы плоское движение маятника происходило в плоскости ($x$ -- $y$). Введем обозначения.
\begin{itemize}
\item $\nu$ --- частота вынуждающих вертикальных гармонических колебаний подвеса,
\item $a$ --- амплитуда вынужденных колебаний,
\item $\omega_0 = \sqrt{g/l}$ --- собственная частота колебаний математического маятника,
\item $g$ --- ускорение свободного падения,
\item $l$ --- длина легкого стержня,
\item $m$ --- масса грузика.
\end{itemize}

\begin{figure}[h]
\begin{center}
\includegraphics[angle=0,width=0.3\columnwidth]{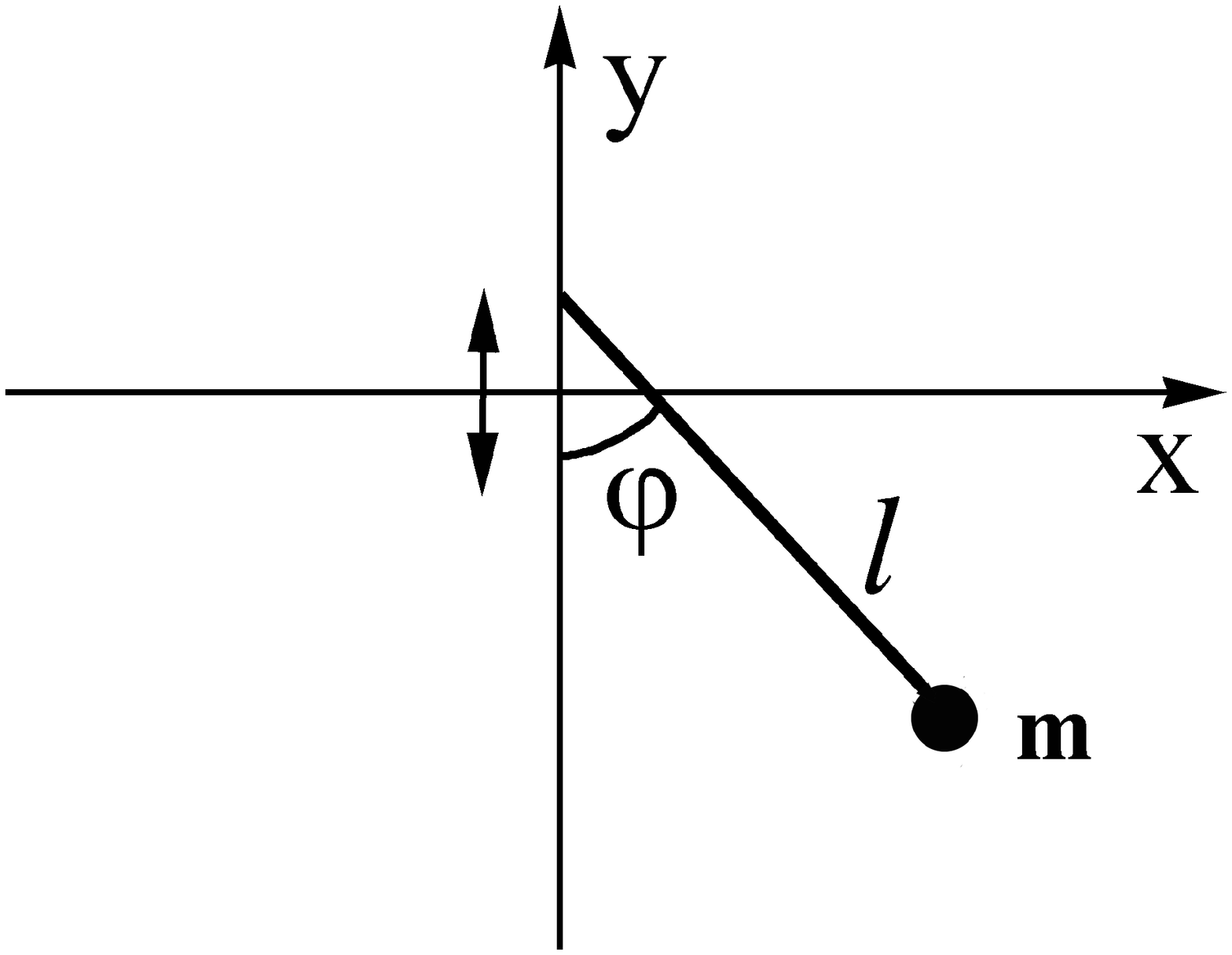}
\end{center}
\caption{Схема маятника.}
\label{FigScheme}
\end{figure}

Если угол между стержнем и осью $y$ обозначить как $\varphi$ (см. Рис.~\ref{FigScheme}), то зависимость координат грузика от времени запишется следующими формулами:
\begin{eqnarray}
\left\{
\begin{array}{lll}
x &=& l \sin \varphi\\
y &=& - l \cos \varphi - a \cos \nu t
\end{array} \right.
\label{Eq:xy(phi)}
\end{eqnarray}

\section{Энергия маятника}

Потенциальная энергия маятника в поле тяжести задается положением грузика по вертикали $y$, как
\begin{eqnarray}
E_{POT} = - m g (l \cos \varphi + a \cos \nu t)\;.
\label{U}
\end{eqnarray}

Для кинетической энергии имеем
\begin{eqnarray}
E_{KIN}
= \frac{m}{2}(\dot x^2 + \dot y^2)
&=& \frac{m}{2}((l \cos \varphi ~\dot \varphi)^2
+(l \sin \varphi ~\dot \varphi + a \nu \sin \nu t)^2)
\nonumber
=\\
&=&
\frac{m l^2 }{2} \dot \varphi^2
+ m a l \nu ~\sin \nu t~ \sin \varphi ~\dot\varphi
+ \frac{m a^2 \nu^2}{2} \sin^2 \nu t
\label{T}
\end{eqnarray}

Полная энергия дается суммой кинетической и потенцальной энергий
\begin{eqnarray}
E = E_{KIN} + E_{POT}\;,
\label{E}
\end{eqnarray}
а лагранжиан системы их разностью
\begin{eqnarray}
L = E_{KIN} - E_{POT}
=
\frac{m l^2 }{2} \dot \varphi^2
+ m a l \nu ~\sin \nu t~ \sin \varphi ~\dot\varphi
+ \frac{m a^2 \nu^2}{2} \sin^2 \nu t
+
m g (l \cos \varphi + a \cos \nu t)\;.
\label{L}
\end{eqnarray}

\begin{figure}
\begin{center}
\includegraphics[angle=0,width=0.4\columnwidth]{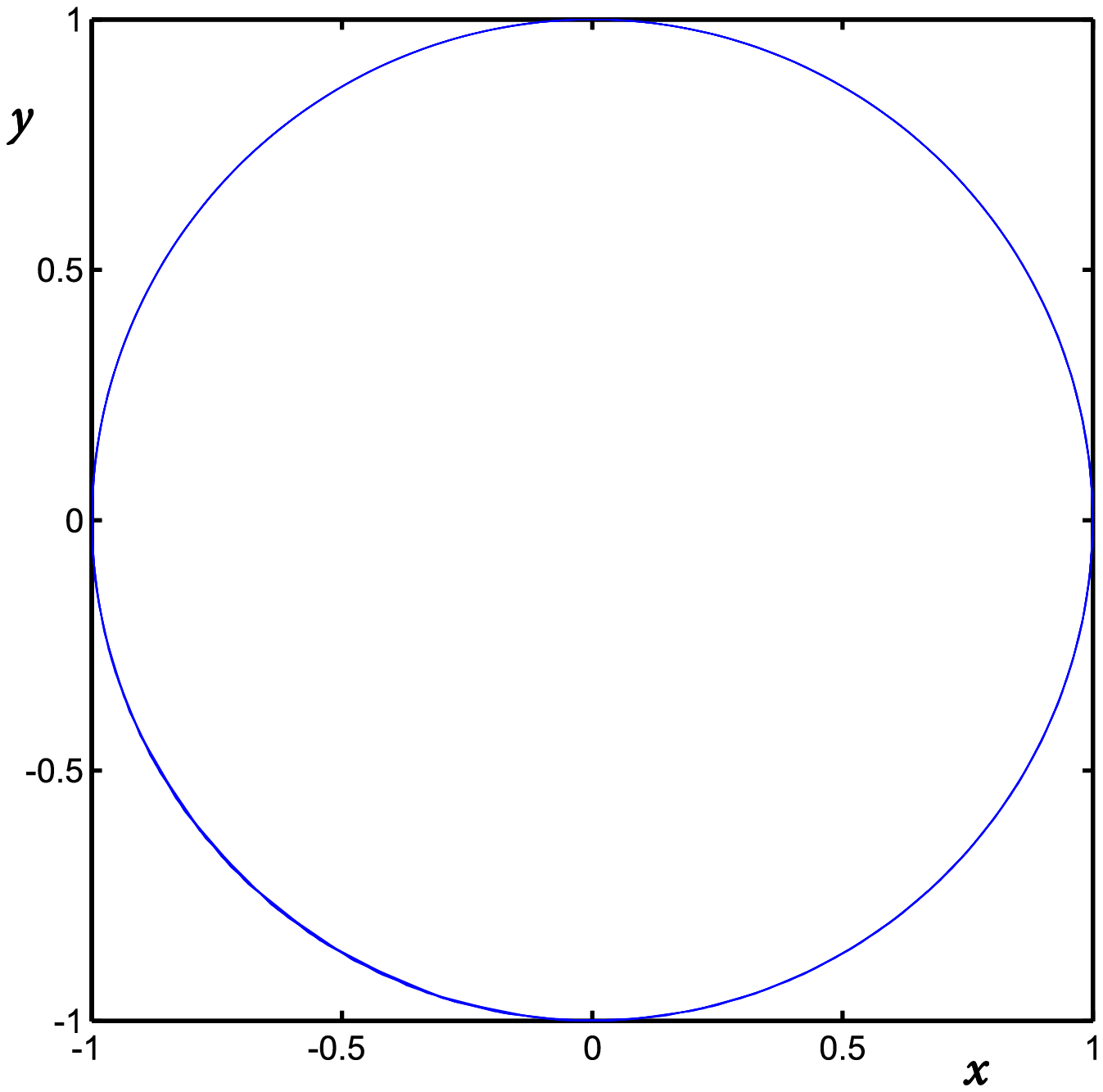}
\includegraphics[angle=0,width=0.4\columnwidth]{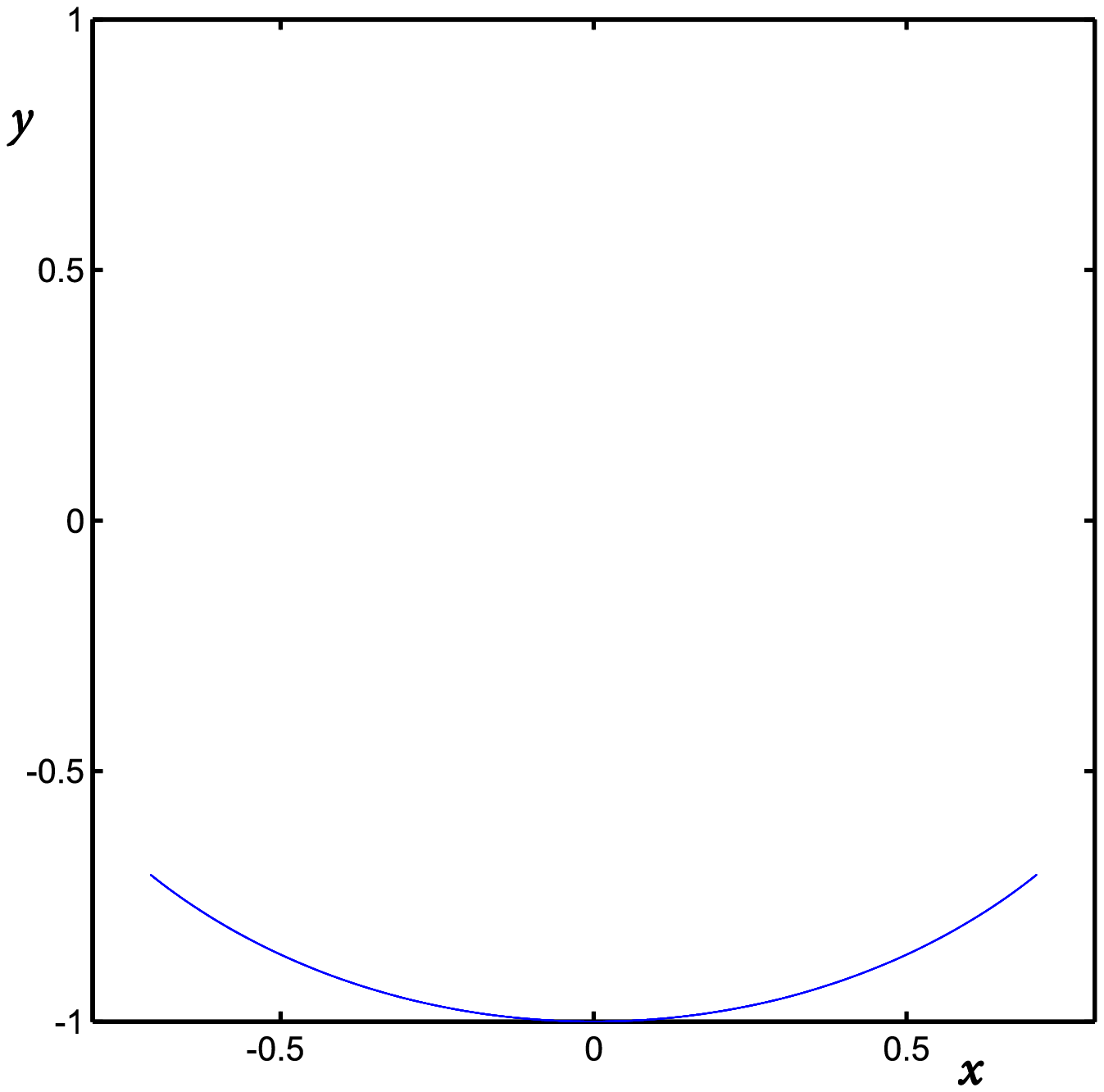}
\end{center}
\caption{Траектория математического маятника. Показаны два характерных случая: (а) $E \ge mgl$ (б) $E < mgl$.}
\label{FigMath}
\end{figure}

\section{Уравнение движения.}

Уравнения движения такой системы удовлетворяют уравнениям Эйлера --- Лагранжа. Отсюда найдем уравнение для зависимости фазы маятника $\varphi$ от времени и таким образом опроеделим положение грузика по формуле~(\ref{Eq:xy(phi)}). Уравнение Эйлера --- Лагранжа для фазы маятника выглядит следующим образом:
\begin{eqnarray}
\frac{d}{dt} \frac{\partial L}{\partial \dot \varphi} = \frac{\partial L}{\partial \varphi}\;.
\end{eqnarray}

Используя формулу (\ref{L}) вычислим те производные, которые используются в этом выражении:
\begin{eqnarray}
\frac{\partial L}{\partial \dot \varphi}
=
m l^2 \dot \varphi
+ m a l \nu ~\sin \nu t~ \sin \varphi
\label{deriv}
\end{eqnarray}
\begin{eqnarray}
\frac{\partial L}{\partial \varphi}
=
m a l \nu ~\sin \nu t~ \cos \varphi ~ \dot \varphi
- m g l \sin \varphi\;.
\end{eqnarray}

Взяв полную производную по времени от (\ref{deriv}) и подставив в (\ref{L}) получим:
\begin{eqnarray}
m l^2 \ddot \varphi
+ m a l \nu^2 ~\cos\nu t~ \sin \varphi
+ m a l \nu ~\sin \nu t~ \cos \varphi ~\dot\varphi
=
m a l \nu ~\sin \nu t~ \cos \varphi ~ \dot \varphi
- m g l \sin \varphi\;.
\end{eqnarray}

Сократив одинаковые слагаемые и поделив правую и левую части на $m l^2$ получим искомое дифференциальное уравнение, описывающие эволюцию фазы маятника
\begin{eqnarray}
\ddot \varphi
=
- \frac{a \nu^2}{l}  ~\cos\nu t~ \sin \varphi
- \frac{g}{l}~ \sin \varphi\;.
\label{diff}
\end{eqnarray}
Отметим, что уравнение (\ref{diff}) нелинейно из-за наличия в нем множителя $\sin\varphi$.

\begin{figure}
\begin{center}
\includegraphics[angle=0,width=0.4\columnwidth]{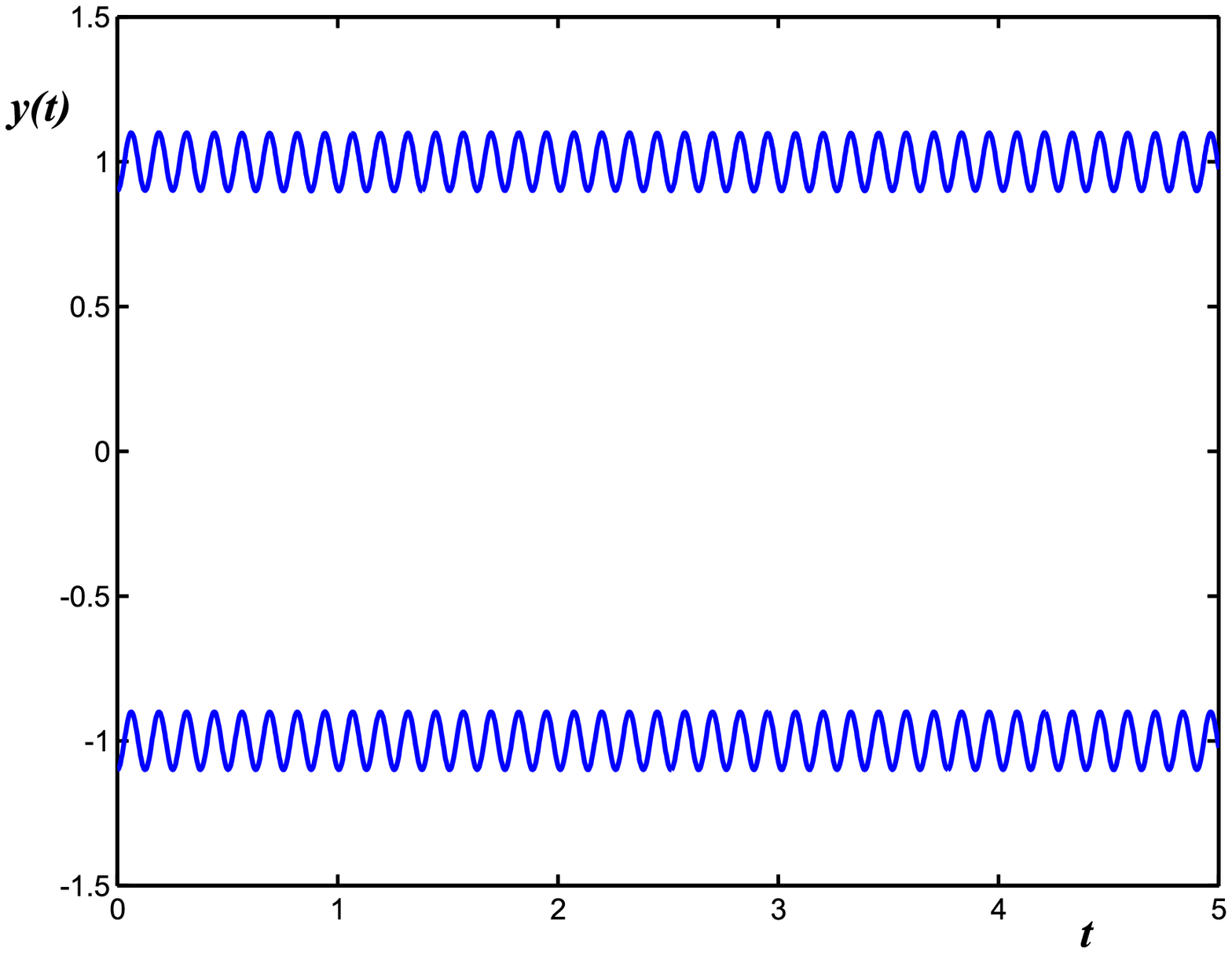}
\includegraphics[angle=0,width=0.4\columnwidth]{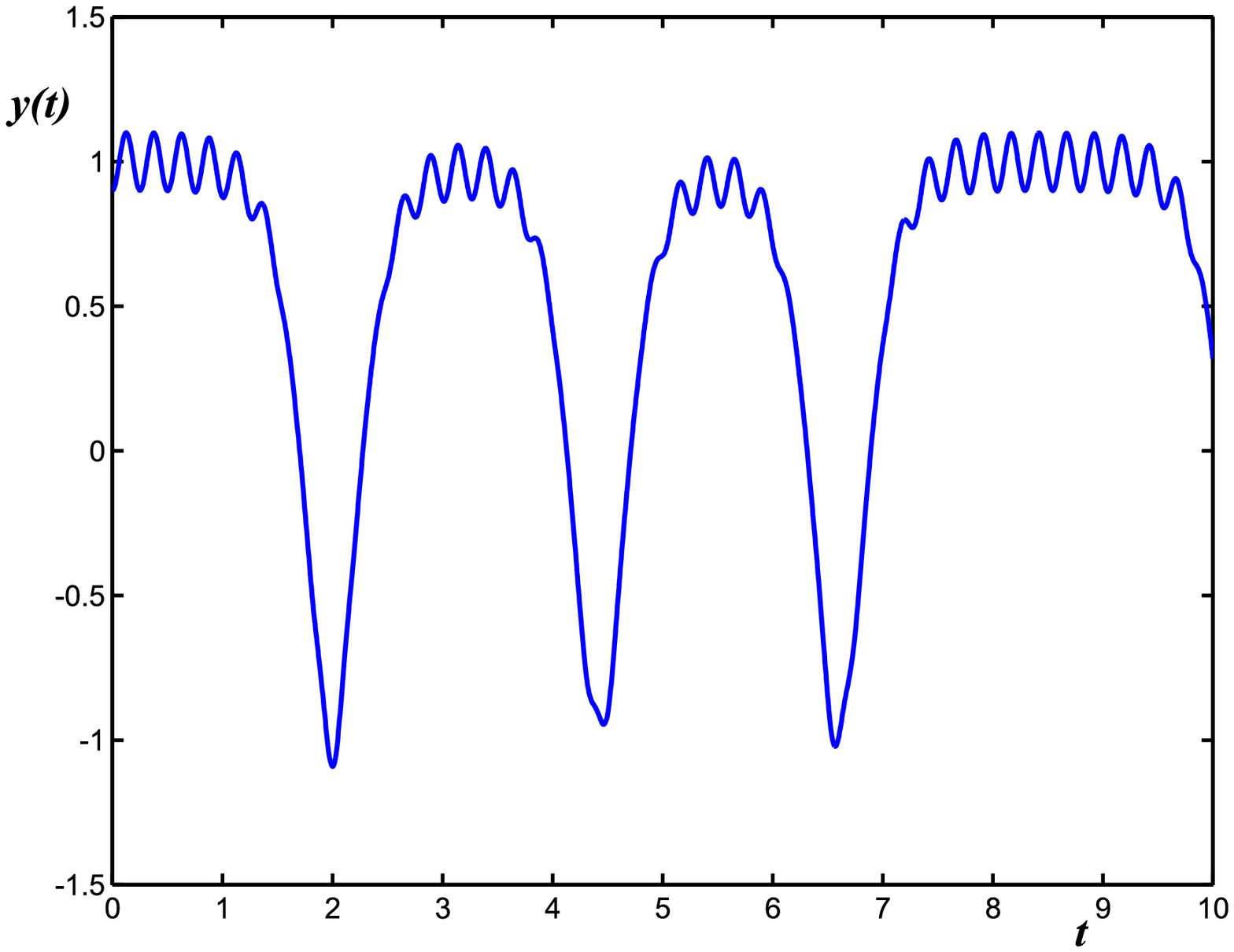}
\end{center}
\caption{Зависимость координаты $y$ от времени для маятника Капицы, $a=0.1$м, $\dot\varphi=50$ рад/сек. (а) $\nu = 50$ Hz начальная фаза $\varphi=0$ для нижнего положение маятника и $\varphi=\pi$ для верхнего положения (б) $\nu = 25$ Hz, $\varphi=\pi$.
Остальные параметры маятника указаны в тексте.}
\label{FigTime}
\end{figure}

\section{Интегрирование уравнения движения.}

Приведем уравнение движения (\ref{diff}) к удобной для интегрирования форме. Для этого обозначим производную от фазы по времени обозначим как $\psi \equiv \dot \varphi$. Тогда данное дифференциальное уравнение, содержащее вторую производную, сведется к системе дифференциальных уравнений, содержащих только первые производные.
\begin{eqnarray}
\left\{
\begin{array}{rcl}
  \dot \varphi &=& \psi\\
  \dot \psi &=& - (\frac{a \nu^2}{l}  ~\cos\nu t~ + \frac{g}{l}~ ) \sin \varphi\\
\end{array}
\right.
\end{eqnarray}

Эта система интегрировалась явным методом Рунге-Кутты.

\begin{figure}
\begin{center}
\includegraphics[angle=0,width=0.32\columnwidth]{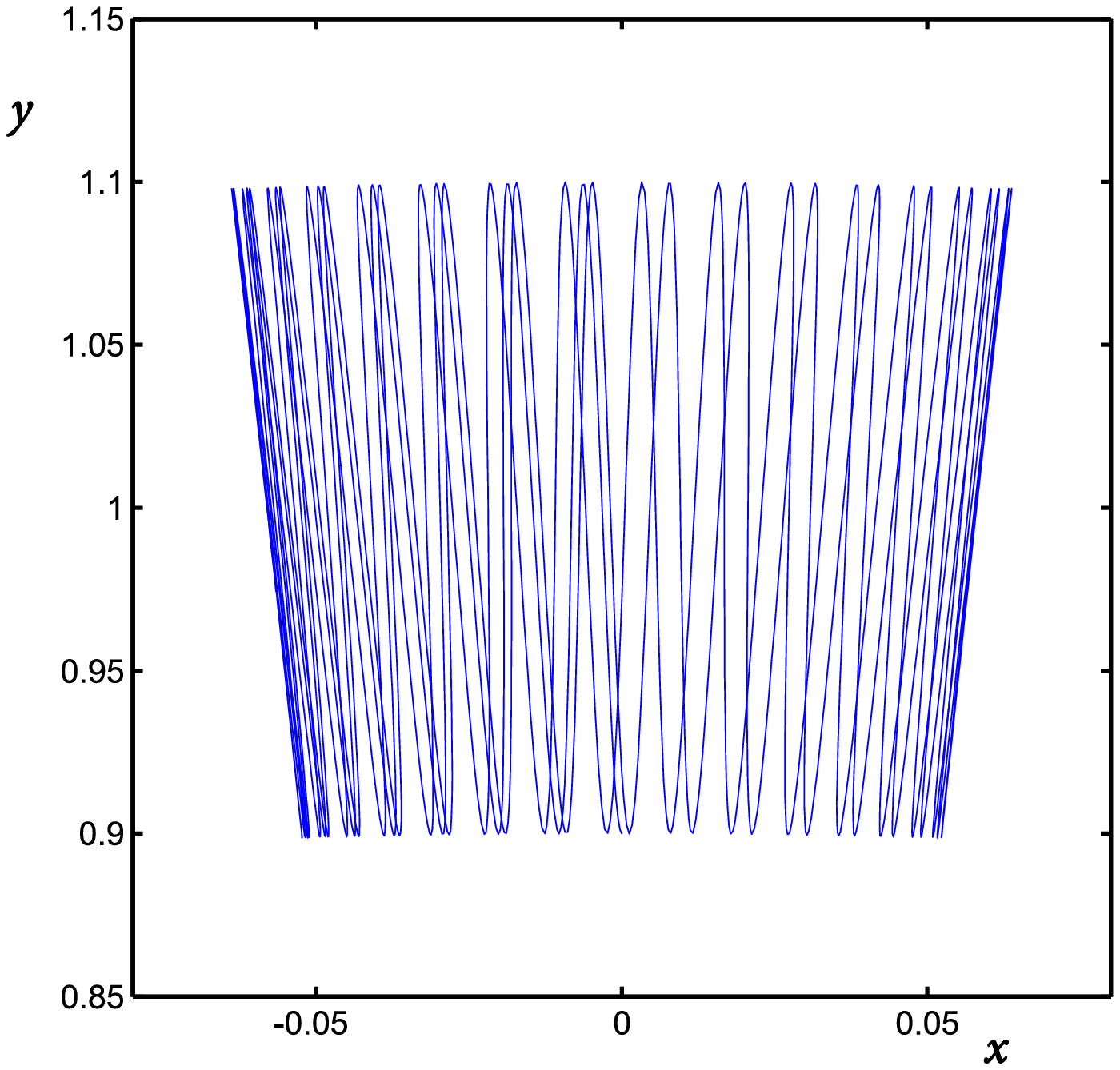}
\includegraphics[angle=0,width=0.32\columnwidth]{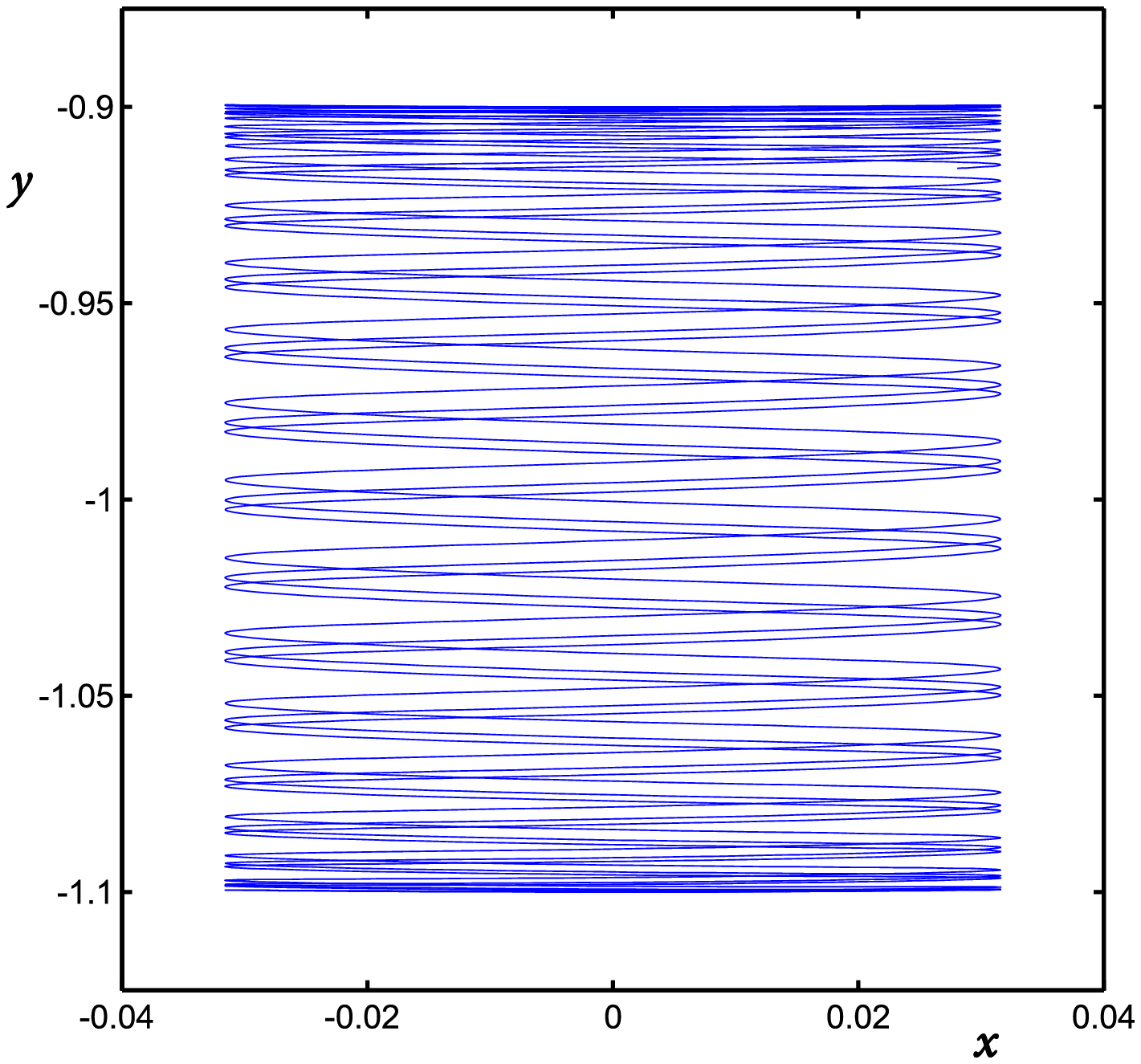}
\includegraphics[angle=0,width=0.32\columnwidth]{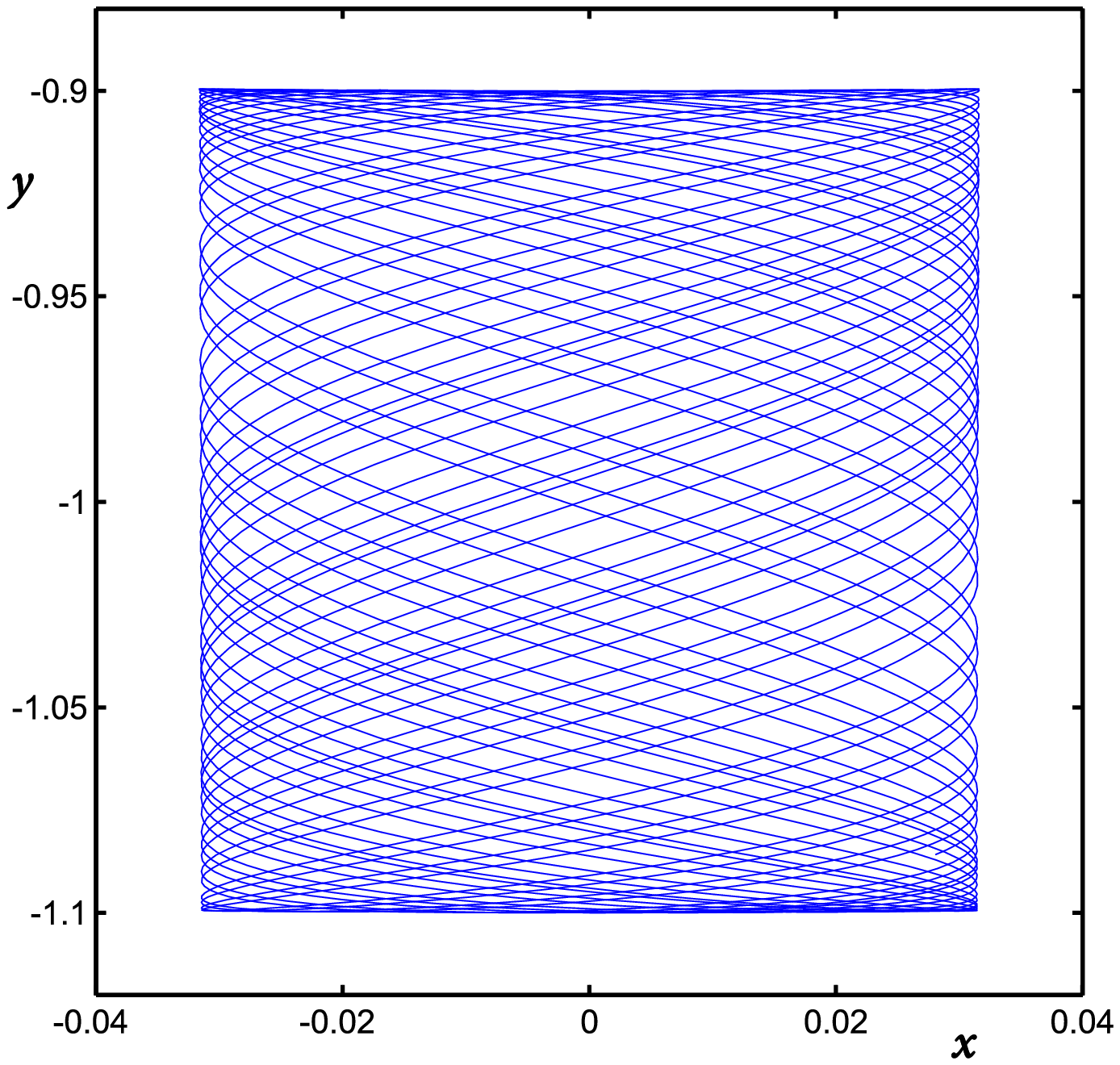}
\end{center}
\caption {Портрет системы в координатном пространстве для маятника Капицы в
(а) в локальном (верхнем) минимуме, (б-в) глобальном (нижнем) минимуме. Вынуждающая частота $\nu$ много больше (а),(б) или порядка (в) собственной частоты маятника $\omega_0$.}
\label{Fig5}
\end{figure}

\section{Положения равновесия.}

Модель маятника Капицы является более общей, чем модель математического маятника. Последняя получается в предельном случае $a = 0$. Фазовый портрет математического маятника хорошо известен. На координатной плоскости  это просто окружность $x^2+y^2 = l^2 = const$. Если в начальный момент времени энергия маятника была больше, чем максимум потенциальной энергии $E > mgl$, то траектория будет замкнутой и циклической, как на Рис.~\ref{FigMath}a. Если же энергия маятника была меньше $E < mgl$, то он будет совершать периодические колебания около единственной устойчивой точки равновесия (см. Рис.~\ref{FigMath}б) с наименьшим значением потенциальной энергии $x = 0, y = -l$. В случае математического маятника полная энергии системы не меняется.

Теперь рассмотрим случай $a \ne 0$. Система более не является замкнутой и ее полная энергия может изменяться. Если при этом, частота вынуждающих колебаний $\nu$ много больше частоты собственных колебаний $\omega_0$, то такой случай можно проанализировать аналитически. Оказывается \cite{Kapitza1,Kapitza2}, что если ввести эффективный потенциал, в котором движется маятник (медленно относительно частоты $\nu$), то этот потенциал имеет два локальных минимума --- один, как и раньше в нижней точке $(0,-l)$, а другой в верхней точке $(0,l)$. То есть точка $(0,l)$ абсолютно неустойчивого равновесия для математического маятника, может оказаться точкой устойчивого равновесия для маятника Капицы.

\begin{itemize}
\item
длина $l$ = 1~м
\item
масса маятника $m$ = 1~кг
\item
амплитуда колебаний подвеса $a$ = 0,1~м
\item
частота вынуждающих колебаний $\nu$ = 50~рад/сек
\item
значение производной по времени от фазы в начальный момент
$\dot\varphi$~=~0,1~рад/cек
\item
ускорение свободного падения $g$ = 10 м/сек$^2$
\end{itemize}

При таких значениях параметров у маятника недостаточно энергии чтобы уйти из локального минимума, однако при частоте колебаний подвеса равной $25$~рад/сек реализуется случай с уходом, показанный на Рис.~\ref{FigTime}б.

\section{Случаи малой вынуждающей частоты и большой амплитуды}

Известны аналитические решения такой задачи при больших значениях $\nu \gg \omega_0$ и малых величинах $a \ll l$. Однако, при отказе от этих ограничений, математическое рассмотрение этой задачи становится слишком сложным. Практически единственным способом решения в общем случае является численное интегрирование уравнения (\ref{diff}). Рассмотрим, как выглядит портрет системы в координатной плоскости. Характерная картина для больших значений частот $\nu$ изображена на Рис.~\ref{Fig5}а. Параметры маятника Капицы такие же,  как и на Рис.~\ref{FigTime}а. Маятник находится в локальном минимуме. Здесь частота $\nu$ гармонических колебаний подвеса является большой по сравнению с собственной частотой математического маятника $\omega_0$, путь грузика имеет выраженные колебания по вертикальному направлению.

Обратная ситуация, когда частота вынуждающих колебаний меньше $\omega_0$ изображена на Рис.~\ref{Fig5}б (в данном случае $\nu = 0,1$, остальные параметры такие же, как и на Рис.~\ref{FigTime}). При таких значениях параметров колебания происходят преимущественно в горизонтальном направлении.

Траектории для промежуточного случая $\nu = 1$ представлены на Рис.~\ref{Fig5}в. Здесь частоты этих двух колебаний соизмеримы, поэтому прямые участки пути располагаются под наклоном.

\begin{figure}
\begin{center}
\includegraphics[angle=0,width=0.4\columnwidth]{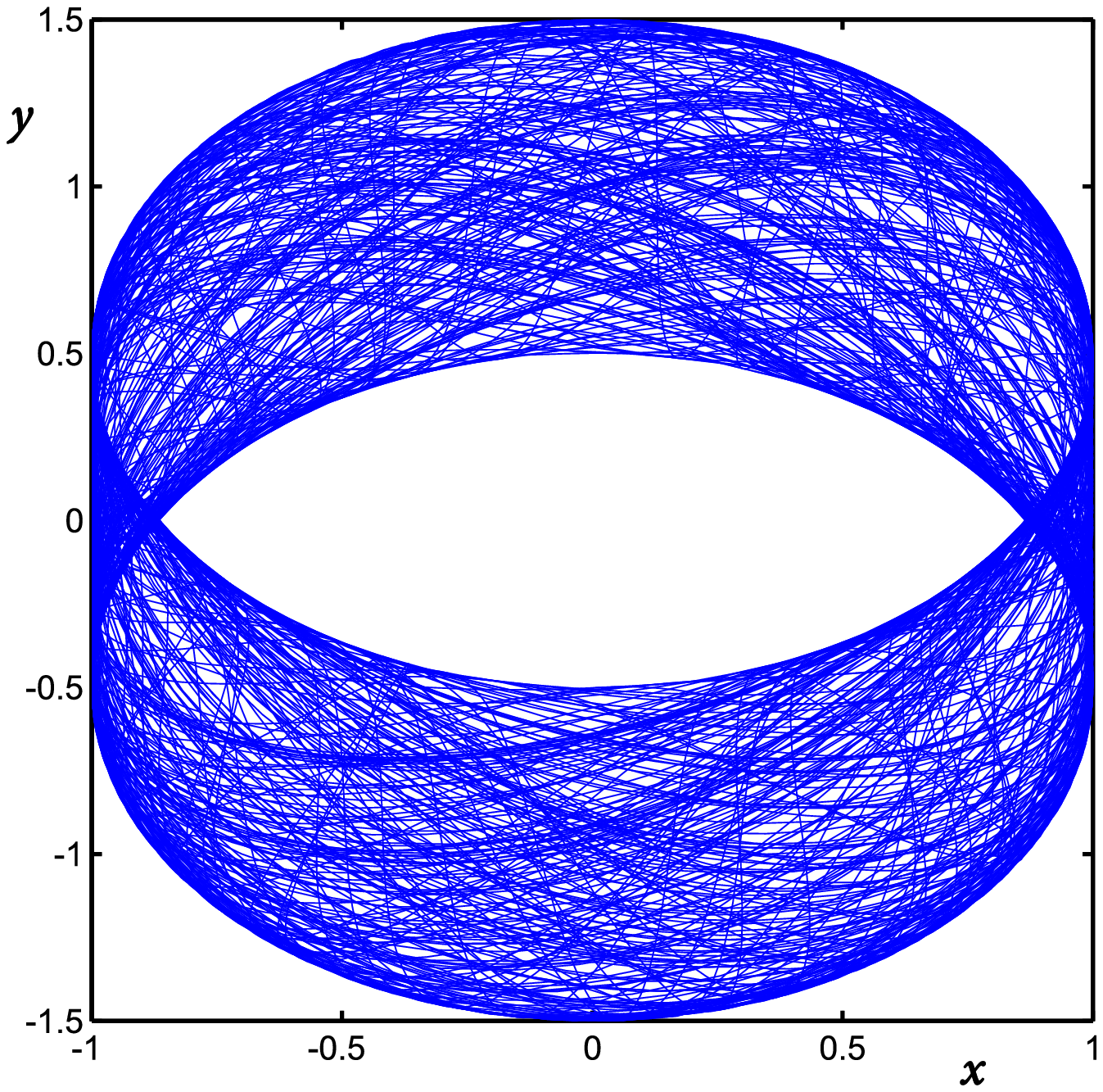}
\includegraphics[angle=0,width=0.4\columnwidth]{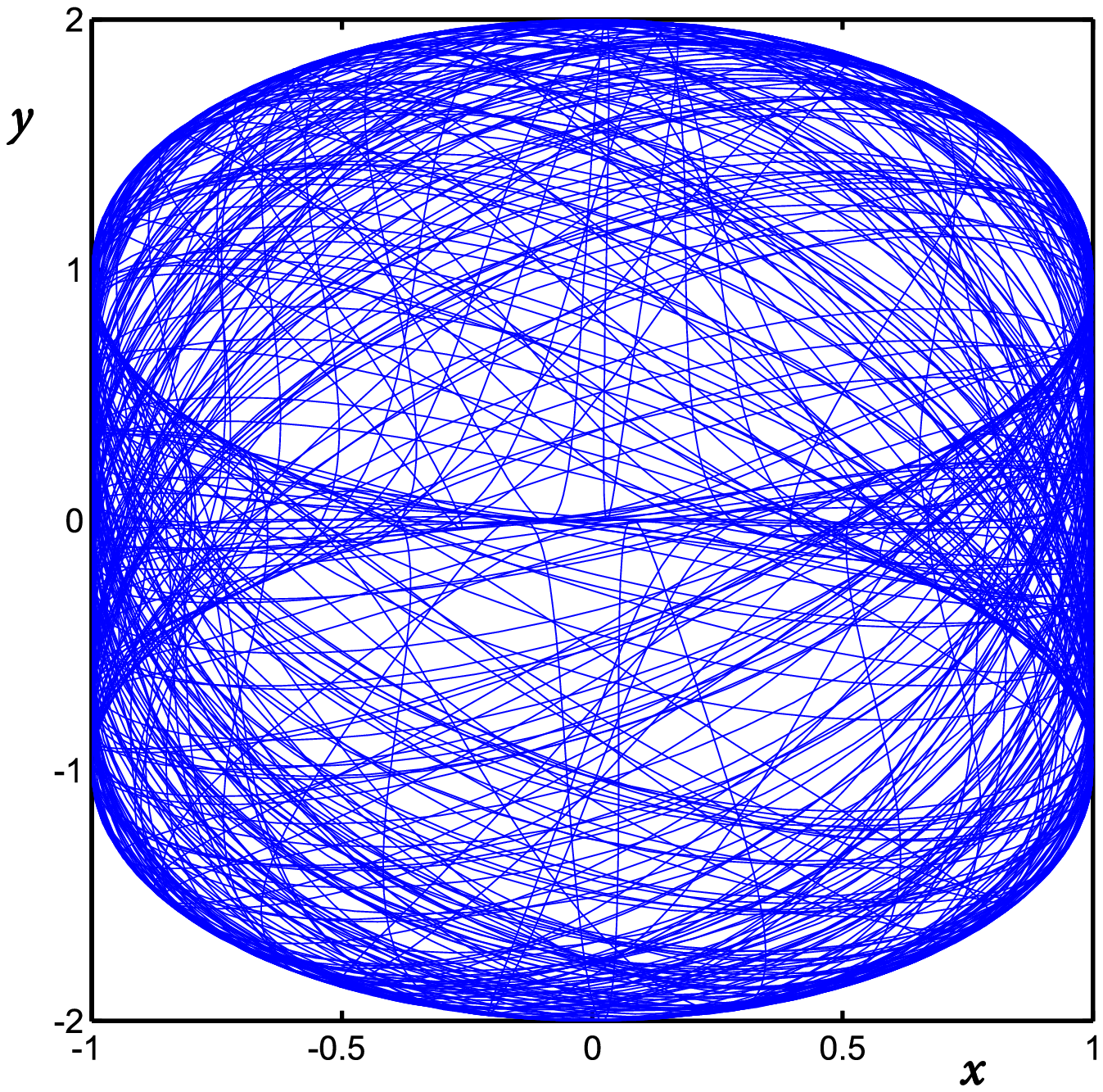}
\end{center}
\caption{Портрет системы в координатном пространстве для маятника Капицы при большой амплитуде вынуждающих колебаний. Параметры маятника приведены в тексте .}
\label{Fig6}
\end{figure}

Интересные фазовые портреты могут быть получены для областей значений параметров, недоступных для аналитического рассмотрения, например в случае большой амплитуды колебания подвеса $a \approx l$. Если увеличить амплитуду вынуждающих колебаний до половины длины маятника $a = l/2$, то получится картина аналогичная той, которая изображена на Рис.~\ref{Fig6}а (у этого маятника частота $\nu = 35$ Hz, угол начального отклонения $\varphi=\pi/2$). При дальнейшем увеличении амплитуды $a$ (начиная от значения $a = l$), все внутреннее пространство начинает <<замазываться>> полностью, т.е., если ранее не все внутренние точки координатного пространства были доступны, то теперь система может побывать в любой точке. Это хорошо видно из Рис.~\ref{Fig6}б ($a = l, \nu = 10$ Hz, $\varphi=\pi/4$). Очевидно, что дальнейшее увеличение длины $a$ принципиально более не изменит картину.

\section{Кинетическая и потенциальная энергии.}

\begin{figure}
\begin{center}
\includegraphics[angle=0,width=0.4\columnwidth]{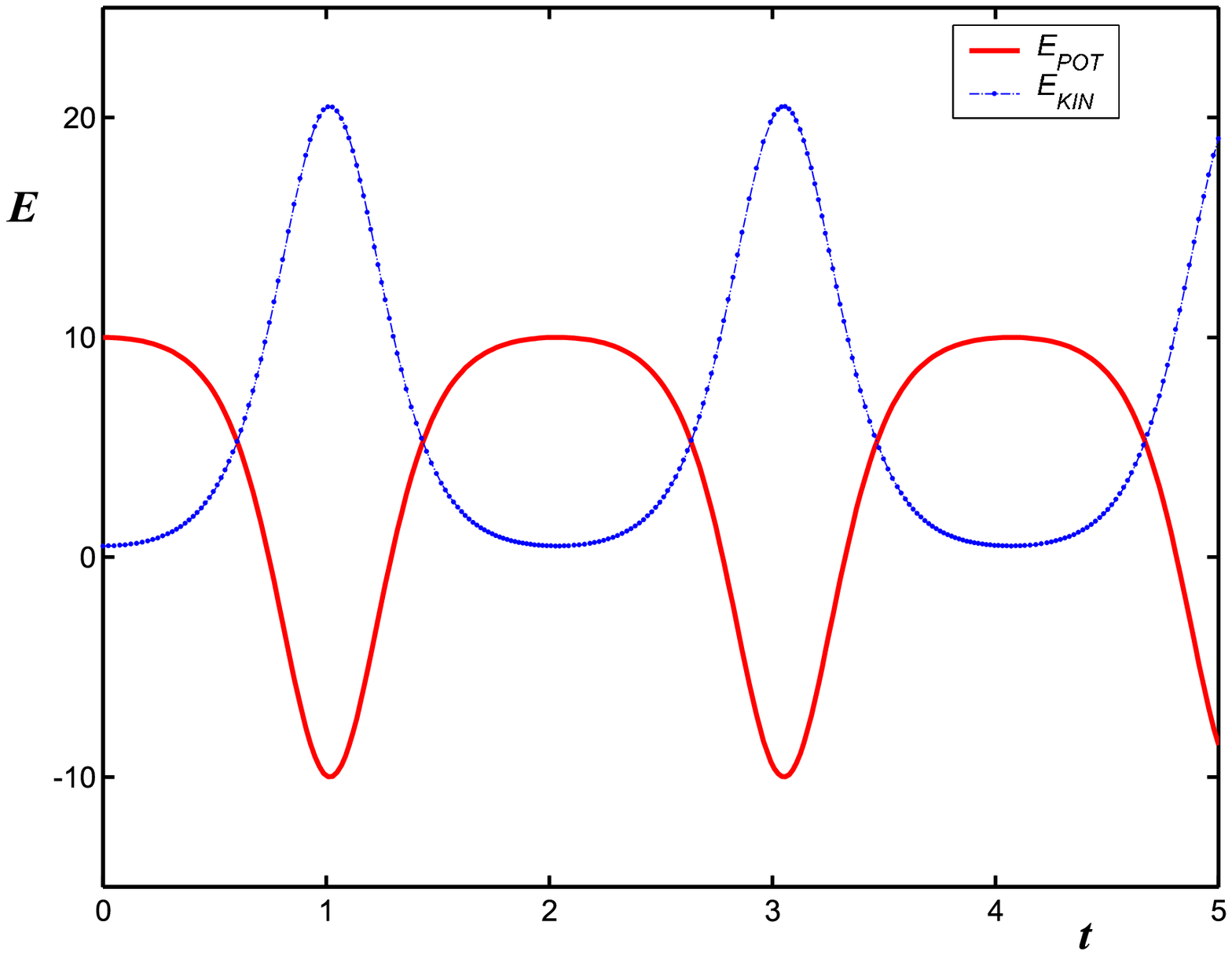}
\includegraphics[angle=0,width=0.4\columnwidth]{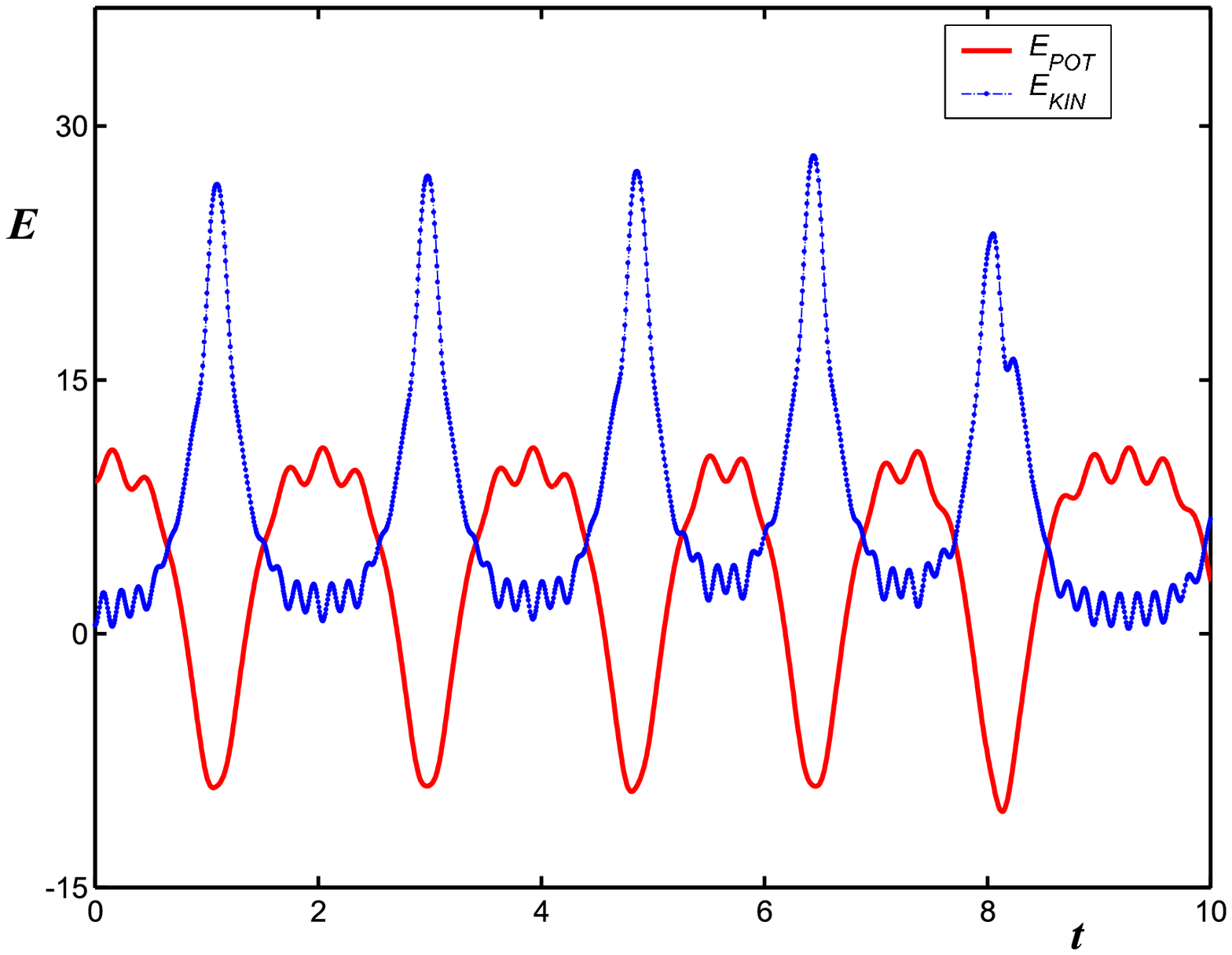}
\end{center}
\caption{Характерные зависимости потенциальной энергии $E_{POT}$ (сплошная линия) и кинетической энергии $E_{KIN}$ (штрих-пунктирная линия) от времени для (а) математического маятника, (б) маятника Капицы. }
\label{Fig7}
\end{figure}

Для математического маятника полная энергия является сохраняющейся величиной, поэтому кинетическая энергия $E_{POT}$ и потенциальная энергия $E_{KIN}$ на графике их зависимости от времени $t$ (Рис.~\ref{Fig7}а) симметричны относительно горизонтальной прямой. Для малых амплитуд колебаний (колебания происходят у глобального минимума, и маятник никогда не делает полный поворот). Из теоремы вириала следует, что средняя кинетическая и потенциальная энергии равны. В этом случае горизонтальная прямая, относительно которой имеется симметрия $E_{KIN}$ и $E_{POT}$, соответствует половине полной энергии (см. Рис.~\ref{Fig7}а).
Если добавить колебания подвеса, $a\ne 0$, то полная энергия больше не будет сохраняться и симметрия больше наблюдаться не будет. Такой случай для маятника с параметрами $a = 0,1~l;~\nu = 20$ изображен на  Рис.~\ref{Fig6}б. Видно, что кинетическая энергия является более чувствительной к вынуждающим колебаниям, чем потенциальная. Отметим, что потенциальная энергия $E_{POT} = mgy$ ограничена как сверху, так и снизу $-l-a<E_{POT}<l+a$, в то время как кинетическая энергия ограничена только снизу $E_{KIN}\ge 0$. При больших значениях частоты $\nu$, кинетическая энергия может быть много больше потенциальной.

\section{Замечания о точности численного решения.}

Численное интегрирование дифференциальных уравнений обладает большим преимуществом перед аналитическим исследованием. А именно, численный подход позволяет отречься от рассмотрения частных, более узких случаев и дает возможность рассматривать свойства системы в большом диапазоне параметров. Однако, за все приходится платить, и машинное интегрирование не является свободным от недостатков. Главным недостатком является возможность накопления большой вычислительной ошибки. В ряде случаев эта ошибка может даже полностью поглотить правильное решение, и полученная картина не будет иметь ничего общего с реальностью.

\begin{figure}
\begin{center}
\includegraphics[angle=0,width=0.4\columnwidth]{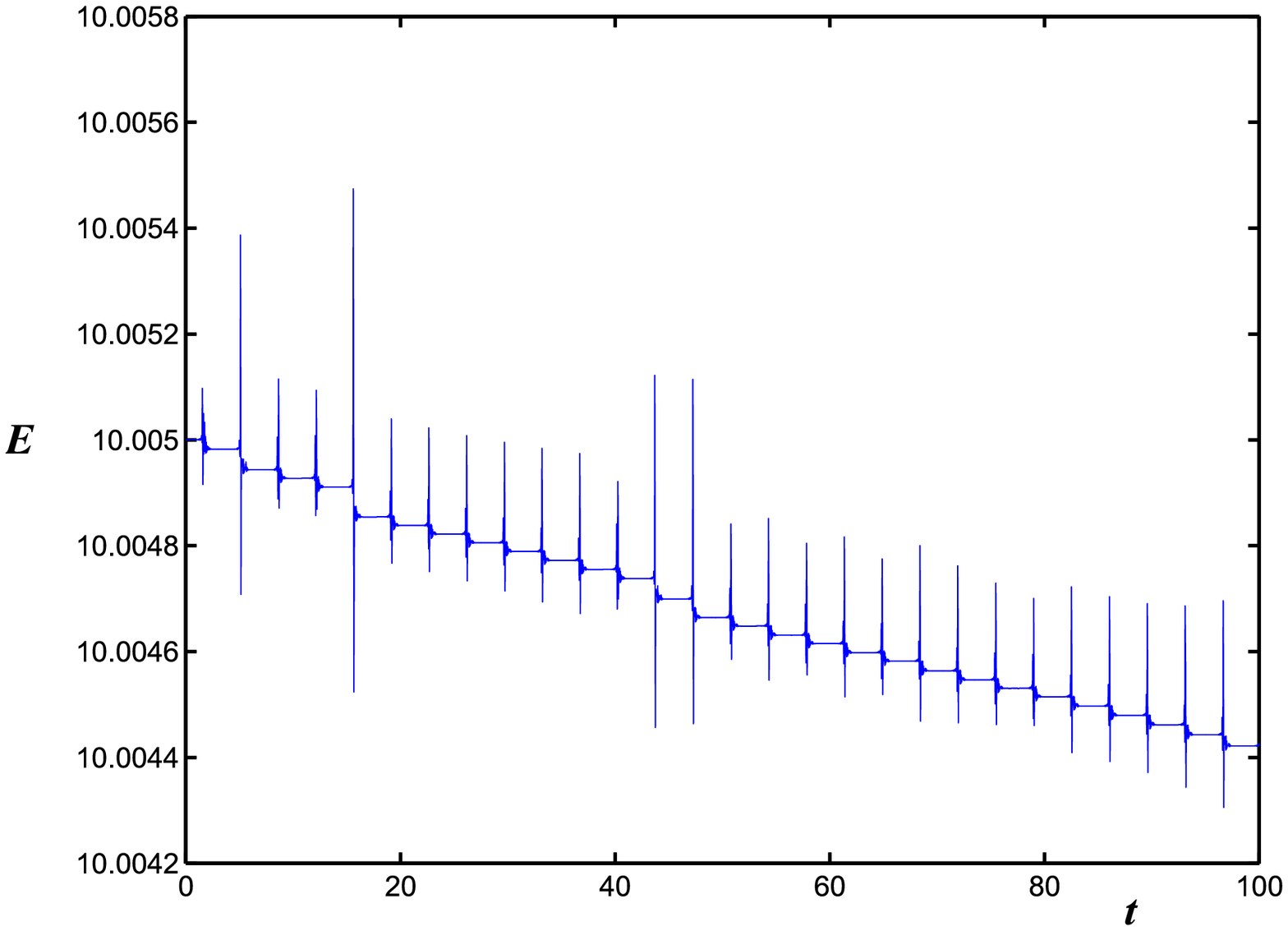}
\end{center}
\caption{Характерный вид зависимость полной энергии математического маятника при численном моделировании.}
\label{Fig8}
\end{figure}

Проиллюстрируем наличие возрастающей ошибки следующим образом: рассмотрим математический маятник. Его полная энергия должна быть величиной постоянной. Промоделируем его изложенным выше способом. Полученные результаты представлены на Рис.~\ref{Fig8}а, как видно полная энергия в этом компьютерном эксперименте уменьшается со временем. В данном случае точкой, дающей наибольший вклад в погрешность, является точка прохождения минимума. На координатной плоскости вместо окружности получается спираль, медленно сходящаяся к центру. Таким образом, нужно быть крайне осторожным при численном моделировании и для получения достоверных результатов необходимо точно знать, откуда возникает погрешность и как ее можно уменьшить.

\section{Интересные факты}

\begin{itemize}
\item
Как отмечал П.Л. Капица, маятниковые часы на вибрирующем основании всегда спешат.

\item
В коридоре института физических проблем стояла работающая модель маятника Капицы,  и любой желающий мог воотчию убедиться, как при ее включении маятник поднимался и оставался в вертикальном положении.

\item
Метод разделения переменных был разработан П.Л.Капицей во время работы над высокочастотным генератором <<ниготроном>>, названным так по месту исследования у себя на даче на Николиной Горе. Для того чтобы не было проблем с <<секретностью>> при публикации метода, П. Л. Капица придумывает простую физическую модель, к которой был бы применим этот метод. Таким образом, появляются статьи \cite{Kapitza1,Kapitza2} про маятник с вибрирующим подвесом.

\item
В измененном варианте П. Л. Капица предлагал решить задачу поступающим к нему в аспирантуру. Требовалась указать условие устойчивости акробата на доске, положенной на цилиндр, лежащий на боку. Ожидаемый ответ был, что если акробат начинал быстро переступать ногами, то его положение становилось устойчивым.

\item
При ходьбе устойчивость тела увеличивается в несколько раз по сравнению с устойчивостью при стоянии. Этот биомеханический феномен до настоящего времени не изучен. Существует гипотеза, которая объясняет устойчивость тела при ходьбе колебательными движениями центра голеностопного сустава. Тело человека представляется с позиции перевернутого маятника с центром в области голеностопных суставов, который приобретает устойчивость в вертикальном положении, если его центр совершает колебание вверх-вниз с достаточно высокой частотой (маятник Капицы).
\end{itemize}

\section{Заключение.}

Итак, в этой работе рассмотрен маятник Капицы, предельным случаем которого является математический маятник. Записаны выражения для потенциальной и кинетической энергий. Приводится вывод уравнения движения маятника во времени, которое решалось численно явным методом Рунге-Кутты. Численно показано, что маятник Капицы обладает двумя устойчивыми положениями, что находится в согласии с теорией. Так же рассмотрены случаи, для которых точное аналитическое решение не известно. Обсуждаются достоинства и недостатки аналитического и машинного способов решений дифференциальных уравнений.


\newpage
\vspace{3cm}
\center{\bf\Large Numerical study of Kapitza pendulum.}

\center{G.E. Astrakharchik}\\
{\it Departament de F\'{\i}sica i Enginyeria Nuclear, Campus Nord B4-B5,\\
 Universitat Polit\`ecnica de Catalunya, E-08034 Barcelona, Spain}

\center{N.A. Astrakharchik}\\
{\it Lyceum, Troitsk, Moscow region, Russia 142190}

\vspace{1cm}
\parbox{12cm}{
A driven pendulum with vertical oscillations of pendulum support (Kapitza pendulum) possesses a number of unusual properties and is a popular object of both analytical and numerical studies. Although some spectacular results can be obtained, such as the vertical position of the pendulum under certain conditions might become stable, no explicit analytical solution for the pendulum trajectory is known. We carry out a numerical study of Kapitza pendulum for a number of different physical regimes. Comparison is made with the limiting cases where the exact solution is known.}


\begin{thebibliography}{20}

\bibitem{Kapitza1} Капица П.Л. <<Динамическая устойчивость маятника при колеблющейся точке подвеса>> ЖЭТФ, {\bf 21}, 588 (1951).

\bibitem{Kapitza2} Капица П.Л. <<Маятник с вибрирующим подвесом>>, УФН, {\bf 44}, 7 (1951)

\bibitem{Stephenson} A. Stephenson ``On an induced stability'' Phil. Mag. 15, 233 (1908)

\bibitem{LL} Ландау Л.Д. и Лифшиц Е. М. Теоретическая физика. Механика. Том 1. Издательство <<Наука>> (1965).

\bibitem{2} Крайнов В.П. Избранные математические методы в теоретической физике. Издательство МФТИ (1996).

\bibitem{Butikov} Бутиков Е. И. <<Маятник с осциллирующим подвесом (к 60-летию маятника Капицы)>>, учебное пособие.
http://faculty.ifmo.ru/butikov/Russian/ParamPendulum.pdf


\bibitem{java} Визуализация в реальном времени движений маятника Капицы доступна в интернете на сайтах
http://www.myphysicslab.com/beta/Inverted-pendulum.html
и
http://faculty.ifmo.ru/butikov/Nonlinear/index.html
Параметры маятника могут быть выбраны произвольно и вводятся вручную.
\end{thebibliography}
\end{document}